\documentclass[conference]{IEEEtran}
\IEEEoverridecommandlockouts
\usepackage{cite}
\usepackage{amsmath,amssymb,amsfonts}
\usepackage{algorithmic}
\usepackage{graphicx}
\usepackage{textcomp}
\usepackage{xcolor}
\usepackage{multirow}
\usepackage{mathtools}
\usepackage{hyperref}
\newcommand{\sign}{\text{sign}}

\usepackage[ruled,vlined]{algorithm2e}
\def\BibTeX{{\rm B\kern-.05em{\sc i\kern-.025em b}\kern-.08em
    T\kern-.1667em\lower.7ex\hbox{E}\kern-.125emX}}
\begin{document}

\title{Robust EEG-based Emotion Recognition Using an Inception and Two-sided Perturbation Model 
\thanks{This work has been partially supported by the National Science Foundation (NSF) under grants CCF-1934962 and DGE-1922591, and by the BRAIN Initiative of the National Institutes of Health through grant U19NS128613. *Corresponding author.}
}

\author{\IEEEauthorblockN{Shadi Sartipi$^{*}$}
\IEEEauthorblockA{\textit{Department of Electrical and Computer Engineering} \\
\textit{University of Rochester}\\
Rochester, USA\\
ssartipi@ur.rochester.edu}
\and
\IEEEauthorblockN{Mujdat Cetin}
\IEEEauthorblockA{\textit{Department of Electrical and Computer Engineering} \\
\textit{Goergen Institute for Data Science}\\
\textit{University of Rochester}\\
Rochester, USA \\
mujdat.cetin@rochester.edu}
}

\maketitle

\begin{abstract}
Automated emotion recognition using electroencephalogram (EEG) signals has gained substantial attention. Although deep learning approaches exhibit strong performance, they often suffer from vulnerabilities to various perturbations, like environmental noise and adversarial attacks. In this paper, we propose an Inception feature generator and two-sided perturbation (INC-TSP) approach to enhance emotion recognition in brain-computer interfaces. INC-TSP integrates the Inception module for EEG data analysis and employs two-sided perturbation (TSP) as a defensive mechanism against input perturbations. TSP introduces worst-case perturbations to the model's weights and inputs, reinforcing the model's elasticity against adversarial attacks. The proposed approach addresses the challenge of maintaining accurate emotion recognition in the presence of input uncertainties. We validate INC-TSP in a subject-independent three-class emotion recognition scenario, demonstrating robust performance.
\end{abstract}

\begin{IEEEkeywords}
Adversarial training,  Brain-computer interfaces, Emotion recognition, Inception module, Weight perturbation.
\end{IEEEkeywords}

\section{Introduction}
\label{sec:intro}
Brain-computer interfaces (BCI) develop a direct pathway between the electrical activities of the human brain and the external environment, bypassing the need for typical nerves and muscles \cite{shih2012brain}. This provides a multidisciplinary framework incorporating psychology, electronics, computers, and neuroscience to interpret the electrical activity of the brain and address health-related challenges to improve life quality \cite{abdulkader2015brain}. Emotions profoundly influence human daily life by impacting physiological activities and decision-making. Automated recognition of emotions can improve human-machine communication \cite{dolan2002emotion}. Among various modalities for emotion recognition, electroencephalography (EEG) stands out as a preferred physiological method for collecting brain signals in BCI research due to its affordability, non-invasiveness, high temporal resolution, and portability \cite{dolan2002emotion}. Various studies have investigated automated emotion recognition by focusing on feature extraction and classification model construction \cite{jenke2014feature}. Some of the widely used features in this context are differential entropy (DE), power spectral density (PSD), and functional connectivity \cite{zheng2015investigating}.

Deep learning (DL) approaches have shown significant performance over traditional machine learning techniques for decoding EEG signals \cite{craik2019deep}. Convolutional neural networks (CNNs) \cite{schirrmeister2017deep} and recurrent neural networks (RNNs) \cite{craik2019deep} are widely used in this area. In \cite{sartipi2023hybrid}, they used the combination of CNN and long short-term memory (LSTM) along with graph-based smoothed EEG data to get the temporal and spatial features for emotion recognition. Popular examples of using CNN-based networks include the DeepCNN and EEGNet proposed in \cite{schirrmeister2017deep} and \cite{lawhern2018eegnet}, respectively.

For various BCI applications, most studies focus on improving the performance of DL approaches. However, one drawback of DL models is that they are often susceptible to precisely crafted minor alterations in input data, which could be due to environmental noise, individual variations, and adversarial attacks \cite{szegedy2013intriguing, zhang2019vulnerability}. These types of perturbations can lead to significant performance degradation \cite{zhang2019vulnerability}. Few studies explored this limitation. For instance, in \cite{meng2019white} the susceptibility of machine learning algorithms in EEG-based BCIs is explored. A novel loss function is proposed in \cite{liu2021universal} to generate universal adversarial perturbations. Since emotion recognition is utilized in mental health applications and various real-time applications \cite{qayyum2020secure}, the robustness of the model alongside the model's performance should be of concern. To the best of our knowledge, there is currently no study on the robustness of EEG-based emotion recognition.


In this paper, we propose a novel DL approach called Inception feature generator and two-sided perturbation (INC-TSP) to extract effective features from EEG data and learn the model in a way that is robust against adversarial attacks for subject-independent emotion recognition. To achieve this, we integrate the Inception module \cite{szegedy2017inception} with a CNN backbone into the deep architecture. As EEG signals contain oscillatory patterns of varying temporal lengths, the Inception module enables a multiscale analysis of the input data. Two-sided perturbation (TSP), which forms an outer maximization problem \cite{wu2020adversarial}, serves as our defensive mechanism. TSP applies worst-case perturbations to the weights and inputs. We evaluate the proposed approach for three-class emotion recognition. Our main contributions can be summarized as follows:
\begin{itemize}
    
    \item We develop an INC-based deep feature generator that performs multiscale analysis of spatial, temporal, and spectral characteristics of EEG data. 
    \item A novel learning approach is used for robust EEG-based emotion recognition against input perturbations. 
\end{itemize}
\begin{table}[t]
\centering
\caption{Inception-based feature generator. Conv is $2$D CNN. Parameters: Conv (number of filters; filter size). Maxpool (kernel size; stride). Dropout (dropout rate).}
\begin{tabular}{lc}
Block& Details \\
\hline
\multirow{1}{*}{C1}& Conv ($64$; $5$), Maxpool ($4$; $4$), Dropout ($0.3$)\\
\hline
\multirow{4}{*}{Inception}&Conv ($32$, $1$)\\
 &Conv ($96$;$1$), Conv ($128$;  $3$)\\
 &Conv ($128$;$3$), Conv ($32$;  $5$)\\
 &Maxpool ($3$;$1$), Conv ($32$;  $5$)\\
\hline
\multirow{1}{*}{C2}& Conv ($256$; $5$), Maxpool ($4$; $4$), Dropout ($0.3$)\\
\hline\\
\end{tabular}
\label{table:inc}
\end{table}
\vspace{-2 pt}
\begin{algorithm}[t]\label{A:algorithm1}
  \SetAlgoLined
  \textbf{Input:} Input data ($x_i,~y_i$), mini-batch size $\hat{m}$, network $f_{\theta}$, PGD iteration $T$, learning rate $\alpha$, TSP step size $\eta_2$, constraint $\epsilon$.\\
 \textbf{Output:} Robust network $f_{\theta}$ \\
\textbf{repeat}\\
    \For{$i=1,\cdots,\hat{m}$}{
    $x_i^{\prime}\leftarrow x_i+\delta,~\text{where}~\delta \sim \text{Uniform}(-\epsilon,+\epsilon)$\\
    \eIf{attack==PGD}
    {
    \For{$t=1,\cdots,$\text{T}}{
    Update $x_i^{\prime}$ based on ~\eqref{eq:eq4}
    }
    }{
    \If{attack==FGSM}
    {
    Update $x_i^{\prime}$ based on ~\eqref{eq:eq3}
    }
    }
    $v\leftarrow v+\eta_{2}\nabla_{v}(\mathcal{L}(f_{\theta+v}(x_i^{\prime}),y_i))$\\
    $v\leftarrow \gamma \frac{v}{\| v\|}\| \theta \|$\\
    }
    $\theta \leftarrow (\theta +v)-\alpha \nabla_{\theta+v} \frac {1}{m}\mathcal{L}(f_{\theta+v}(x_i^{\prime},y_i))-v$
    
\textbf{until training converge}
  \caption{TSP}
\end{algorithm}
\section{Method}
\noindent {\bf Inception-based Feature Generator:} The process of feature extraction plays a pivotal role in EEG emotion recognition studies. To achieve this, we draw inspiration from image classification techniques and employ the Inception module \cite{szegedy2017inception} to extract features. The designed inception-based (INC) feature generator architecture comprises a combination of multiple convolutional layers, a pooling layer, and an activation function.

Let the input data be denoted as $X \in \mathbb{R}^{n\times{}c\times{}t}$, where $n$, $c$, and $t$ are the number of frequency subbands, EEG channels, and temporal length of the data, respectively. Table~\ref{table:inc} shows specific configurations and architectural choices for three different CNN blocks. The feature extraction process involves the initial application of a convolutional layer, C1, to acquire shallow features. Subsequently, the inception module, with different convolutional kernels, is employed to capture deeper features. The various kernel sizes allow the model to capture diverse information in both the temporal and spatial domains. The output features from the parallel convolutional layers are then merged using a concatenation layer. The concatenated features pass through another convolutional layer, C2, followed by pooling and dropout layers in the final step. ReLU activation function and batch normalization are applied in CNN layers. Finally, the features are fed to three fully connected layers with dimensions of $512$, $256$, and $64$.

\noindent {\bf Robust Generalization:}The performance of deep neural networks for EEG-based emotion recognition is often vulnerable to perturbations applied to input data, which could significantly diminish the model's performance across different subjects \cite{zhang2019vulnerability}. These perturbations are also known as adversarial attacks \cite{szegedy2013intriguing}.

To overcome such attacks, we employ adversarial training (AT) with adversarial weight perturbation as a defensive scheme inspired by \cite{wu2020adversarial}. This approach involves the use of two-sided perturbation (TSP), which applies worst-case perturbations to both the input data and the model's weights.
\begin{table*}[t!]
\centering
\caption{Subject-independent robust accuracy and robust F1 score of the proposed INC-TSP.}
\begin{tabular}{lccccccc}
& \multicolumn{2}{c}{PGD-10}&\multicolumn{2}{c}{PGD-20}&\multicolumn{2}{c}{FGSM}\\
Threat Model & R-Accuracy & R-F1-score&R-Accuracy & R-F1-score&R-Accuracy & R-F1-score\\
\hline
$L_2$&$0.91\pm 0.04$& $0.89\pm 0.05$& $0.86\pm 0.07$& $0.85\pm 0.08$& $0.85\pm 0.05$&$0.84\pm 0.05$\\
$L_{\infty}$&$0.82\pm 0.07$ & $0.81\pm 0.07$& $0.81\pm 0.08$& $0.080 \pm 0.08$& $0.84\pm 0.08$&$0.83\pm 0.07$\\
\hline
\end{tabular}
\label{table:atawp}
\end{table*}

\subsubsection{ Adversarial Attack}
To begin, we briefly explain two widely used adversarial attack techniques targeted at computer vision models. Fast gradient sign method (FGSM) \cite{goodfellow2014explaining} is a single-step gradient-based method to find the perturbed example in one step by the amount of $\epsilon$ in a direction specified by the sign of the gradient of the loss function:
\begin{equation}
\label{eq:eq3}
    x_i^{\prime}=x_i+\epsilon . \sign (\nabla_{x_i}\mathcal{L}(f_{\theta}(x_i^{\prime}),y_i))
\end{equation}

Another attack is projected gradient descent (PGD) \cite{madry2017towards} which perturbs the original sample for $T$ iterations with step-size $\eta$. At each iteration, PGD projects the perturbed sample back onto the $\epsilon$-ball at the $t^{th}$ iteration as:
\begin{equation}
\label{eq:eq4}
    x_i^{\prime (t+1)}=\Pi_{\epsilon}(x_i^{\prime (t)}+\eta . \sign (\nabla_{x_i^{\prime}}\mathcal{L}(f_{\theta}(x_i^{\prime (k)}),y_i)))
\end{equation}
where $\Pi$ is the projection operator.

\subsubsection{ Adversarial Defense}
AT directly integrates adversarial examples into the training process to flatten the loss changes with respect to the input via the optimization problem presented below \cite{madry2017towards}:
\begin{equation}\label{eq:EQ1}
    \min_{\theta}\rho(\theta),~\text{where}~\rho(\theta)=\frac{1}{n}\sum_{i=1}^{n} \max_{\delta \leq \epsilon}\mathcal{L}(f_{\theta}(x_i^{\prime}),y_i)
\end{equation}

where $\rho(\theta)$ is the adversarial loss, $n$ is the number of samples, $\delta=\|x_i^{\prime}-x_i\|_{p}$, $\|.\|_{p}$ is the norm (threat model), $x_i^{\prime}$ is the adversarial example within the $\epsilon$-ball centered at original sample $x_i$, $f_{\theta}$ is the deep learning architecture, i.e., INC, with weights $\theta$, $\mathcal{L}$ is the classification loss, and $y_i$ are the true labels. Following \cite{wu2020adversarial}, to incorporate the flatness of the loss change with respect to the weight, we propose solving the following optimization problem:
\begin{equation}
\label{eq:tsp}
    \min_{\theta}\max_{v}\rho(\theta+v)=\min_{\theta}\max_{v}\frac{1}{n}\sum_{i=1}^{n} \max_{\delta \leq \epsilon}\mathcal{L}(f_{\theta+v}(x_i^{\prime}),y_i)
\end{equation}
where $v$ is selected from the feasible region for weight perturbation. Let $\gamma$ be the constraint on weight perturbation size. The weight perturbation on the $l^{th}$ layer with weight $\theta_{l}$ is $\|v_{l}\|\leq \gamma \| \theta_{l}\|$.

\subsubsection{Learning process}
The optimization and learning process for INC-TSP is shown in Algorithm~\ref{A:algorithm1}. Let $\hat{m}$ be the mini-batch size. The input perturbation is applied by choosing the attack type subsequently the weight perturbation is calculated. Finally, the model parameters, $\theta$, are updated via the  Adam optimizer.

\begin{figure}[t]
    \centering
    \begin{minipage}[c]{.14\textwidth}
    \includegraphics[width=\textwidth]{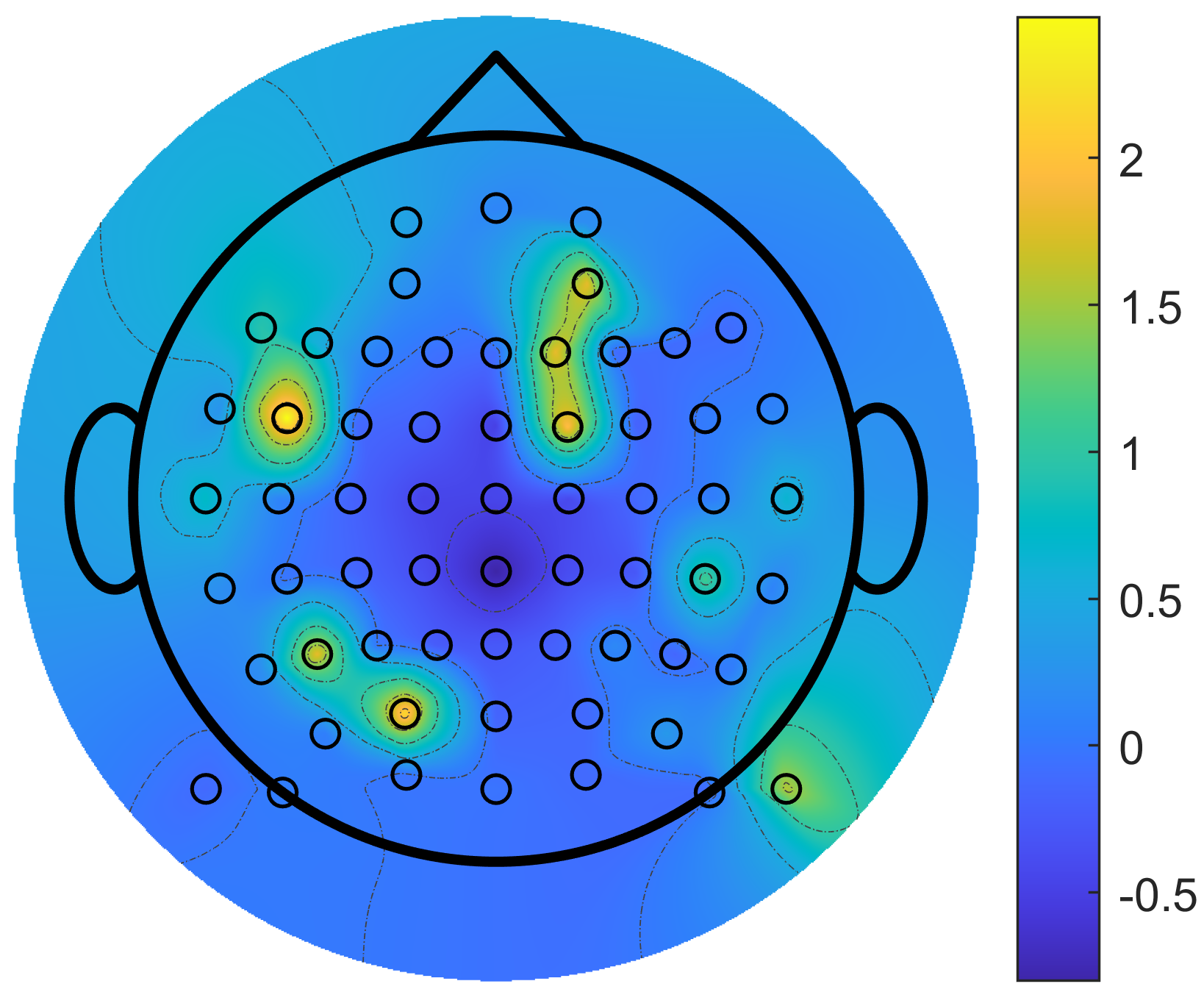}
    \end{minipage}
    \begin{minipage}[c]{.14\textwidth}
    \includegraphics[width=\textwidth]{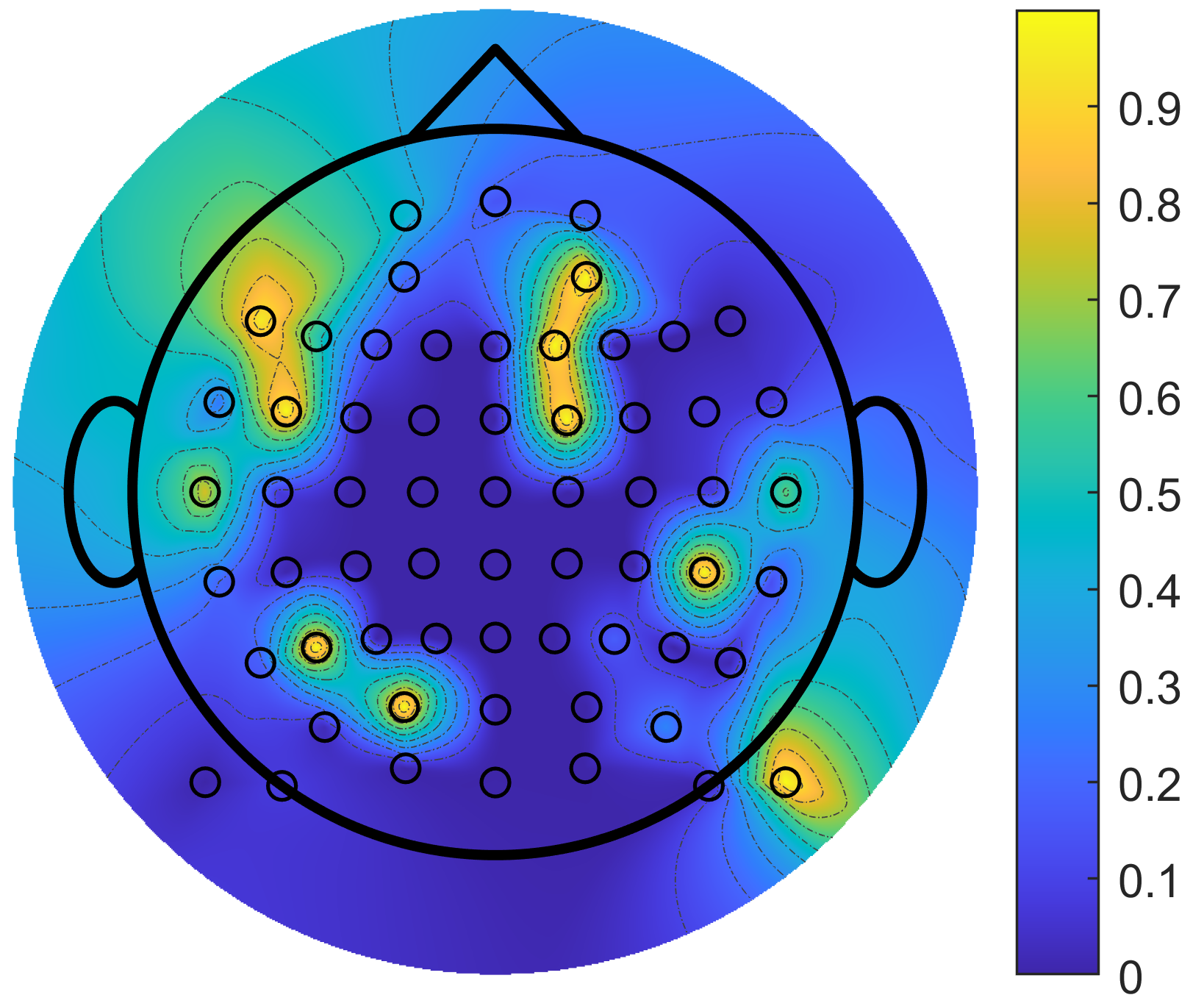}
    \end{minipage}
    \begin{minipage}[c]{.14\textwidth}
    \includegraphics[width=\textwidth]{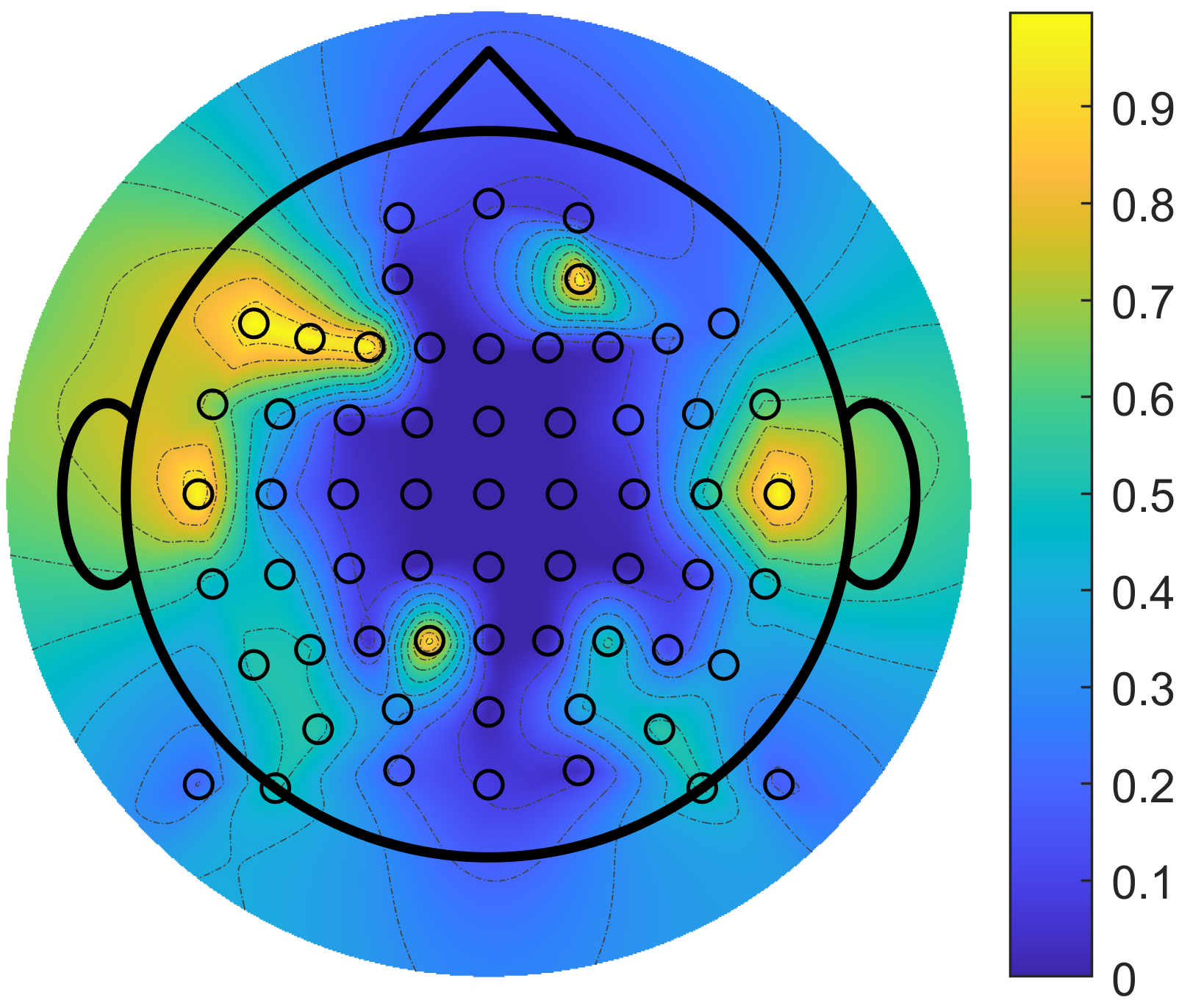}
    \end{minipage}
\begin{minipage}{0.15\textwidth}
 \centering
\text{(a)}
\end{minipage}
\begin{minipage}{0.15\textwidth}
 \centering
\text{(b)}
\end{minipage}
\begin{minipage}{0.15\textwidth}
 \centering
\text{(c)}
\end{minipage}
    \caption{Sample EEG features for (a) original data, (b) perturbed with PGD-10, and (c)  perturbed with PGD-20.}
    \label{fig:samles}
\end{figure}

\section{Experimental Study}
\label{sec:results}
\begin{figure*}[t]
    \centering
    \begin{minipage}[c]{.32\linewidth}
    \includegraphics[width=0.9\textwidth]{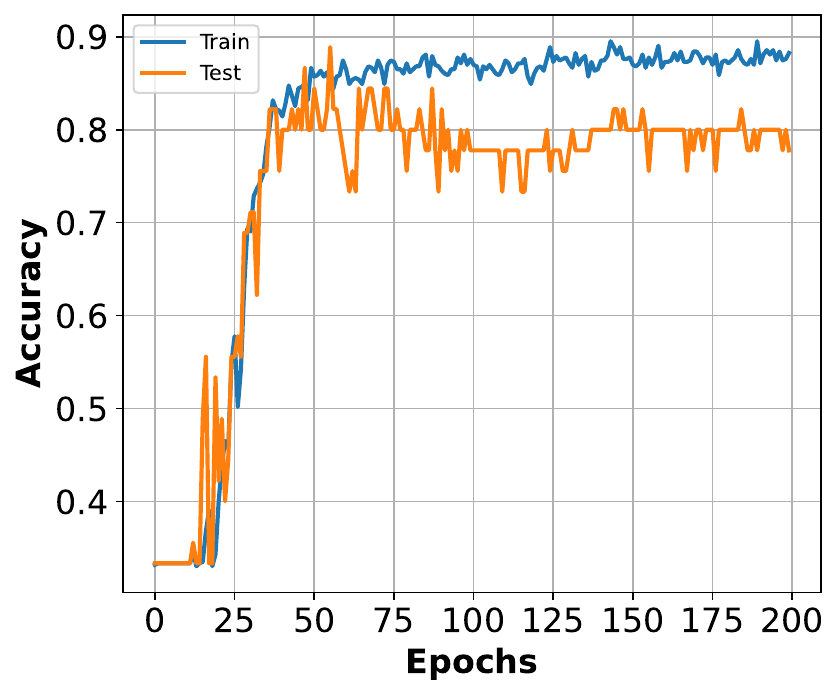}
    \end{minipage}
\hfill
    \begin{minipage}[c]{.32\linewidth}
    \includegraphics[width=0.9\textwidth]{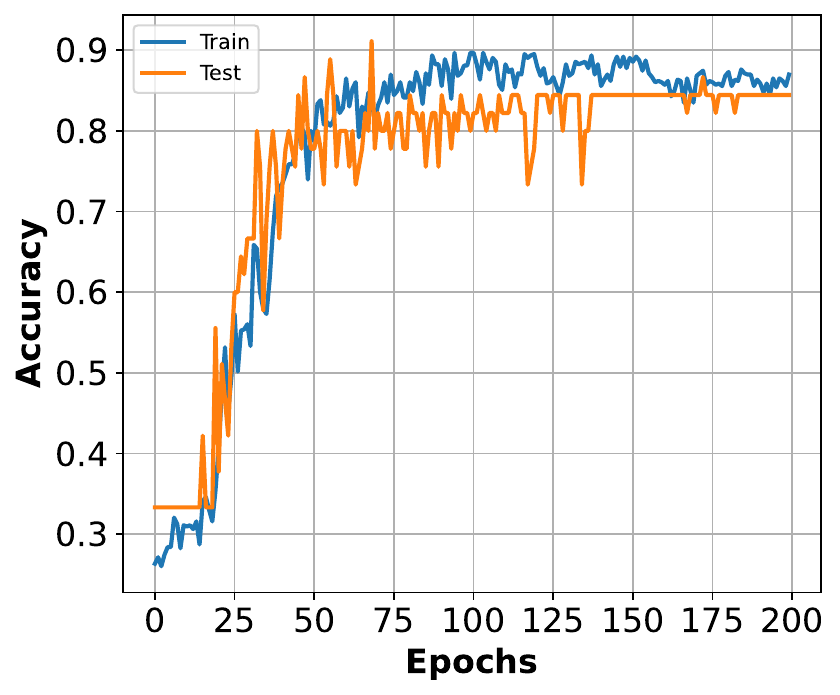}
    \end{minipage}
    \hfill
    \begin{minipage}[c]{.32\linewidth}
    \includegraphics[width=0.9\textwidth]{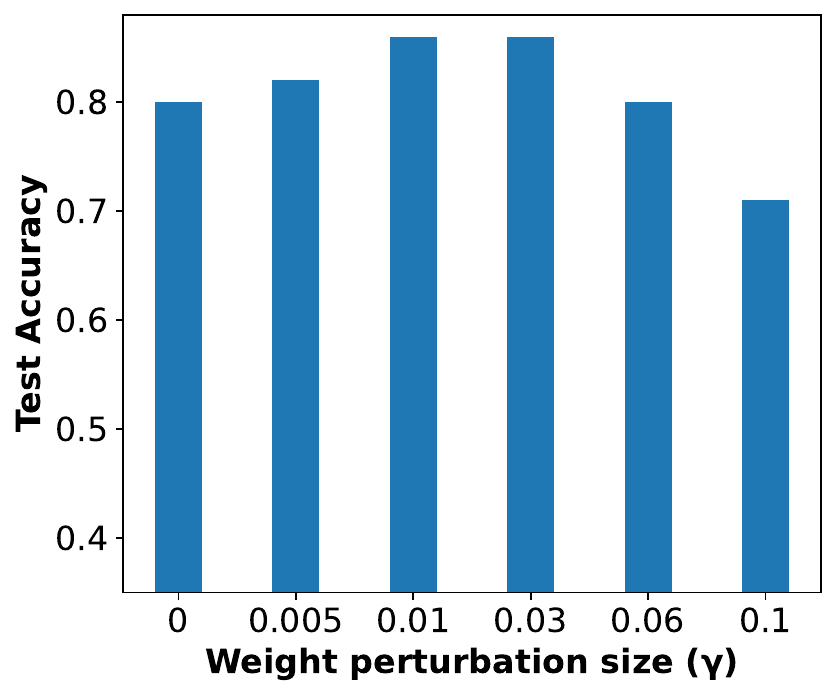}
    \end{minipage}
\begin{minipage}{0.32\textwidth}
 \centering
\text{(a)}
\end{minipage}
\begin{minipage}{0.32\textwidth}
 \centering
\text{(b)}
\end{minipage}
\begin{minipage}{0.32\textwidth}
 \centering
\text{(c)}
\end{minipage}
\vspace{-10 pt}
    \caption{Learning curve of PGD-10 attack (a) INC without defense, (b) INC-TSP, and (c) robustness of INC-TSP as a function of weight perturbation size.}
    \label{fig:originalvsrobust}
\end{figure*} 
\noindent {\bf Dataset:} Here, we use the publicly available SEED dataset \cite{zheng2015investigating} for emotion recognition. The dataset comprises fifteen movie clips designed to elicit three emotions: happiness, sadness, and neutrality. The experiments involved a total of 15 participants, consisting of 8 females and 7 males, and were instructed to immerse themselves in the movie clips to evoke the corresponding emotions. The EEG signals were recorded using the international $10$-$20$ system with $62$ channels. Each trial followed a specific sequence: a 5-second starting hint before the film clip, 4 minutes of the clip as an emotional stimulus, 45 seconds for self-assessment, and a 15-second break. The recorded EEG data were downsampled from $1000$ Hz to $200$ Hz, and a band-pass filter with a frequency range of $0.5$-$70$ Hz was applied. We calculated the differential entropy (DE) features every $1$-second with no overlap in five frequency subbands, namely, delta, theta, alpha, beta, and gamma.
\begin{table}[t]
\centering
\caption{Ablation study for a PGD-10 attack.}
\begin{tabular}{lcc}
Defense & R-Accuracy & Accuracy\\
\hline
TSP&$\textbf{0.91}\pm \textbf{0.04}$ &$\textbf{0.93}\pm \textbf{0.03}$\\
AT&$0.81\pm 0.06$ &$0.88\pm 0.04$\\
Without defense&$0.55\pm 0.08$ &$0.86\pm0.06$\\
\hline
\end{tabular}
\label{table:ablation}
\end{table}
\begin{figure}[t]
    \centering
    \begin{minipage}[c]{.23\textwidth}
    \includegraphics[width=\textwidth]{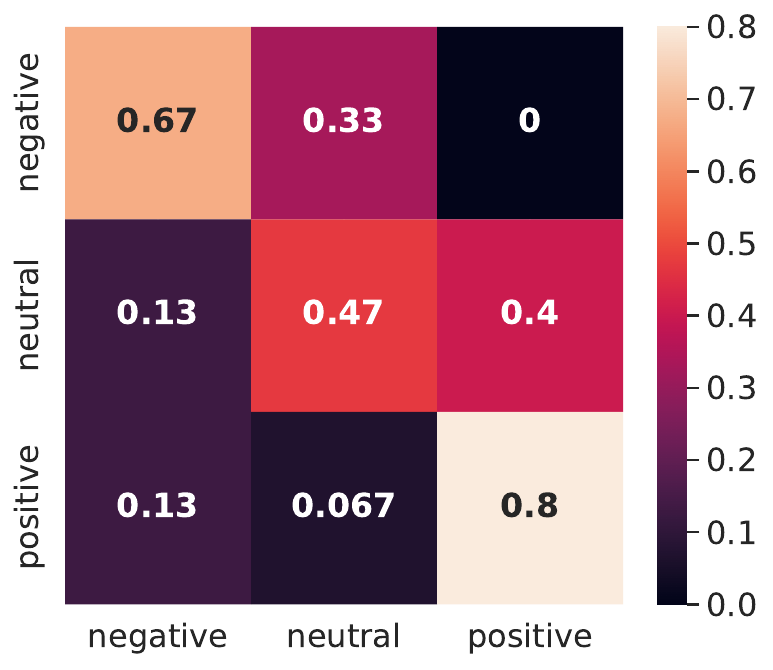}
    \end{minipage}
    \hfill
    \begin{minipage}[c]{.23\textwidth}
    \includegraphics[width=\textwidth]{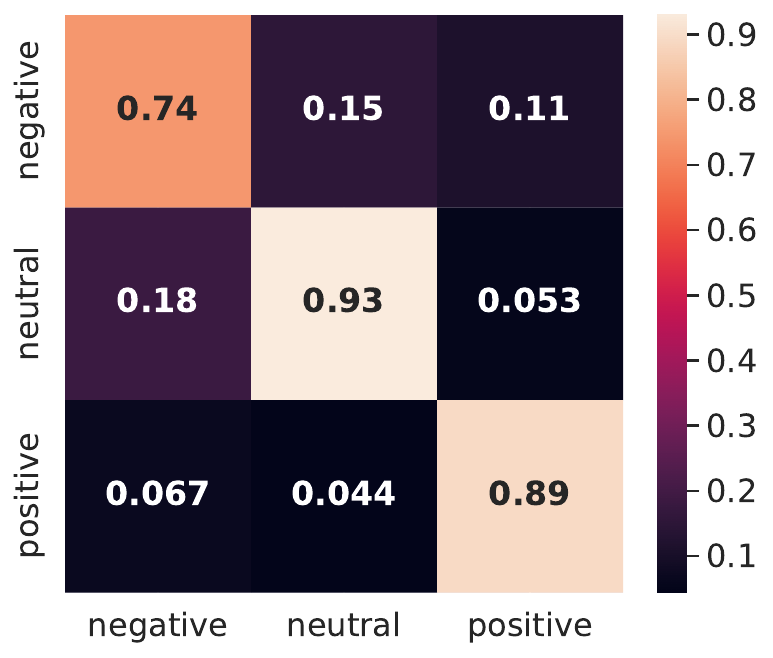}
    \end{minipage}
    \hfill
    \caption{Confusion matrices for a perturbed input with PGD (\textbf{left}) without defense and (\textbf{right}) with TSP.}
    \label{fig:confusion}
\end{figure} 

\begin{table}[t]
\centering
\caption{Comparison with previous works.}
\begin{tabular}{lc}
Study & Accuracy \\
\hline
\textbf{Proposed method}&$\textbf{0.93}\pm \textbf{0.03}$ \\

SDDA \cite{li2022dynamic}&$0.91\pm 0.07$\\
MSFR-GCN \cite{pan2023msfr}&$0.87\pm 0.05$\\
SECT\cite{bai2023sect}&$0.85\pm 0.06$\\
\hline
\end{tabular}
\label{table:comparison}
\end{table}

\noindent {\bf Results:} In this section, we present the performance of the proposed approach. We employed the $L_2$ and $L_{\infty}$ threat models. The value of $\epsilon$ is set to $8/225$, and the PGD step size is set to $15/255$. We evaluated the approach against various adversarial attacks. Samples of the perturbed input with $10$ and $20$ iterations, PGD-10 and PGD-20, are depicted in Fig.~\ref{fig:samles}. The weight perturbation size and learning rate are both set to $0.03$ and $9 \times 10^{-5}$, respectively.

We employed leave-one-subject-out (LOSO) cross validation to assess performance. The input data are normalized by subtracting the mean and dividing by the standard deviation. To evaluate the performance and show the effectiveness of the defense mechanism, we computed the robust accuracy and F1-score (R-Accuracy and R-F1-score). R-Accuracy and R-F1-score present the accuracy and F1-score on the perturbed test data. The average performance across subjects is presented in Table~\ref{table:atawp}. As shown in Table~\ref{table:atawp}, the best robust performance is $0.91 \pm 0.04$ under the $L_2$ threat model and PGD-10 adversarial attacks. Furthermore, these results suggest that INC-TSP consistently maintains R-Accuracy and R-F1-scores, indicating its effectiveness in countering adversarial perturbations and ensuring accurate classification.

Furthermore, we performed the ablation study shown in Table~\ref{table:ablation}. R-Accuracy and Accuracy columns are the performance of each approach on test data with PGD-10 attack and non-perturbed data, respectively. Comparing the results without defense to those with AT \eqref{eq:EQ1} and TSP \eqref{eq:tsp}, the proposed learning process achieves the highest R-Accuracy and accuracy. For detailed performance analysis, we present the confusion matrices for classification using the INC architecture without a defense and with the proposed INC-TSP approach with PGD-20 attack in Fig.~\ref{fig:confusion}. This demonstrates TSP's ability to accurately detect all three emotional states even when the model encounters perturbed inputs. Table~\ref{table:comparison} compares INC-TSP's performance to previous works, demonstrating INC-TSP's superiority. All methods considered here use a LOSO cross-validation process.

\noindent {\bf Effect of TSP on robustness and generalization:} To investigate the generalization of the proposed learning approach, we present the learning curves in the training and test phases with and without TSP (Fig.~\ref{fig:originalvsrobust} (a) and (b)). The major drawback of DL approaches in BCI is the overfitting of the model on the training data and poor performance on new test subjects. As presented in Fig.~\ref{fig:originalvsrobust} (b), the learning curves of INC-TSP for test and training follow each other closely which shows the generalization and robustness of the learned model. Additionally, Fig.~\ref{fig:originalvsrobust} (c) presents the model performance across different weight perturbation sizes. These results show that while the model gets the best performance with $\gamma=0.01~\text{and}~0.03$, the performance does not drop significantly even with large perturbations.

\section{Conclusion}
\label{sec:conclusion}
This study proposed the INC-TSP model as a novel approach to enhance the robustness of emotion recognition in BCIs. By combining the Inception module and TSP as a defensive mechanism, the model demonstrates durability against adversarial attacks and input uncertainties. Results indicate that the INC-TSP model consistently achieves robust accuracy and F1-scores across various threat models and adversarial attack scenarios which shows its efficacy in countering perturbations. The robustness and generalization investigations suggest the use of INC-TSP as a possible defense mechanism in future BCI implementations.

\bibliographystyle{IEEEbib}
\bibliography{refs}

\begin{thebibliography}{10}

\bibitem{shih2012brain}
Jerry~J Shih, Dean~J Krusienski, and Jonathan~R Wolpaw,
\newblock ``Brain-computer interfaces in medicine,''
\newblock {\em Mayo clinic proceedings}, vol. 87, no. 3, pp. 268--279, 2012.

\bibitem{abdulkader2015brain}
Sarah~N Abdulkader, Ayman Atia, and Mostafa-Sami~M Mostafa,
\newblock ``Brain computer interfacing: Applications and challenges,''
\newblock {\em Egyptian Informatics Journal}, vol. 16, no. 2, pp. 213--230, 2015.

\bibitem{dolan2002emotion}
Raymond~J Dolan,
\newblock ``Emotion, cognition, and behavior,''
\newblock {\em science}, vol. 298, no. 5596, pp. 1191--1194, 2002.

\bibitem{jenke2014feature}
Robert Jenke, Angelika Peer, and Martin Buss,
\newblock ``Feature extraction and selection for emotion recognition from {EEG},''
\newblock {\em {IEEE} Trans. Affect. Comput.}, vol. 5, no. 3, pp. 327--339, 2014.

\bibitem{zheng2015investigating}
Wei-Long Zheng and Bao-Liang Lu,
\newblock ``Investigating critical frequency bands and channels for {EEG}-based emotion recognition with deep neural networks,''
\newblock {\em IEEE Transactions on autonomous mental development}, vol. 7, no. 3, pp. 162--175, 2015.

\bibitem{craik2019deep}
Alexander Craik, Yongtian He, and Jose~L Contreras-Vidal,
\newblock ``Deep learning for electroencephalogram ({EEG}) classification tasks: a review,''
\newblock {\em J. Neural Eng.}, vol. 16, no. 3, pp. 031001, 2019.

\bibitem{schirrmeister2017deep}
Robin~Tibor Schirrmeister and {et al.},
\newblock ``Deep learning with convolutional neural networks for {EEG} decoding and visualization,''
\newblock {\em Human brain mapping}, vol. 38, no. 11, pp. 5391--5420, 2017.

\bibitem{sartipi2023hybrid}
Shadi Sartipi, Mastaneh Torkamani-Azar, and Mujdat Cetin,
\newblock ``A hybrid end-to-end spatio-temporal attention neural network with graph-smooth signals for {EEG} emotion recognition,''
\newblock {\em {IEEE} Trans. Cogn. Develop. Syst.}, 2023.

\bibitem{lawhern2018eegnet}
Vernon~J Lawhern and {et al.},
\newblock ``{EEGNet}: a compact convolutional neural network for {EEG}-based brain--computer interfaces,''
\newblock {\em J. Neural Eng.}, vol. 15, no. 5, pp. 056013, 2018.

\bibitem{szegedy2013intriguing}
Christian Szegedy, Wojciech Zaremba, Ilya Sutskever, Joan Bruna, Dumitru Erhan, Ian Goodfellow, and Rob Fergus,
\newblock ``Intriguing properties of neural networks,''
\newblock {\em arXiv preprint arXiv:1312.6199}, 2013.

\bibitem{zhang2019vulnerability}
Xiao Zhang and Dongrui Wu,
\newblock ``On the vulnerability of {CNN} classifiers in {EEG}-based {BCIs},''
\newblock {\em {IEEE} Trans. Neural Syst. Rehabil. Eng.}, vol. 27, no. 5, pp. 814--825, 2019.

\bibitem{meng2019white}
Lubin Meng and {et al.},
\newblock ``White-box target attack for {EEG}-based {BCI} regression problems,''
\newblock in {\em Neural Information Processing: 26th International Conference, ICONIP 2019, Sydney, NSW, Australia, December 12--15, 2019, Proceedings, Part I 26}. Springer, 2019, pp. 476--488.

\bibitem{liu2021universal}
Zihan Liu, Lubin Meng, Xiao Zhang, Weili Fang, and Dongrui Wu,
\newblock ``Universal adversarial perturbations for {CNN} classifiers in {EEG}-based {BCIs},''
\newblock {\em J. Neural Eng.}, vol. 18, no. 4, pp. 0460a4, 2021.

\bibitem{qayyum2020secure}
Adnan Qayyum, Junaid Qadir, Muhammad Bilal, and Ala Al-Fuqaha,
\newblock ``Secure and robust machine learning for healthcare: A survey,''
\newblock {\em IEEE Reviews in Biomedical Engineering}, vol. 14, pp. 156--180, 2020.

\bibitem{szegedy2017inception}
Christian Szegedy, Sergey Ioffe, Vincent Vanhoucke, and Alexander Alemi,
\newblock ``Inception-v4, inception-resnet and the impact of residual connections on learning,''
\newblock {\em Proceedings of the AAAI conference on artificial intelligence}, vol. 31, no. 1, 2017.

\bibitem{wu2020adversarial}
Dongxian Wu, Shu-Tao Xia, and Yisen Wang,
\newblock ``Adversarial weight perturbation helps robust generalization,''
\newblock {\em Advances in Neural Information Processing Systems}, vol. 33, pp. 2958--2969, 2020.

\bibitem{goodfellow2014explaining}
Ian~J Goodfellow, Jonathon Shlens, and Christian Szegedy,
\newblock ``Explaining and harnessing adversarial examples,''
\newblock {\em arXiv preprint arXiv:1412.6572}, 2014.

\bibitem{madry2017towards}
Aleksander Madry, Aleksandar Makelov, Ludwig Schmidt, Dimitris Tsipras, and Adrian Vladu,
\newblock ``Towards deep learning models resistant to adversarial attacks,''
\newblock {\em arXiv preprint arXiv:1706.06083}, 2017.

\bibitem{li2022dynamic}
Zhunan Li and {et al.},
\newblock ``Dynamic domain adaptation for class-aware cross-subject and cross-session {EEG} emotion recognition,''
\newblock {\em {IEEE} J. Biomed. Health Inform.}, vol. 26, no. 12, pp. 5964--5973, 2022.

\bibitem{pan2023msfr}
Deng Pan, Haohao Zheng, Feifan Xu, Yu~Ouyang, Zhe Jia, Chu Wang, and Hong Zeng,
\newblock ``Msfr-gcn: A multi-scale feature reconstruction graph convolutional network for {EEG} emotion and cognition recognition,''
\newblock {\em {IEEE} Trans. Neural Syst. Rehabil. Eng.}, 2023.

\bibitem{bai2023sect}
Zhongli Bai and {et al.},
\newblock ``Sect: A method of shifted {EEG} channel transformer for emotion recognition,''
\newblock {\em {IEEE} J. Biomed. Health Inform.}, 2023.

\end{thebibliography}

\end{document}